\documentclass[twoside,final,journal]{IEEEtran}
\usepackage[latin1]{inputenc}
\usepackage{graphicx}
\usepackage{subfigure}
\usepackage{amsmath}
\usepackage{color}
\usepackage[normalem]{ulem}
\usepackage[latin1]{inputenc}
\usepackage{graphicx}
\usepackage{subfigure}
\usepackage{verbatim}
\usepackage{amsmath}
\usepackage{textcomp}
\usepackage{amssymb}
\usepackage{amsfonts}
\usepackage{multirow}
\usepackage{color}
\usepackage[noadjust]{cite}
\usepackage{bm}
\usepackage{cite}
\usepackage{balance}

\begin{document}
\title{Adaptive Background Compensation of FI-DACs with Application to
  Coherent Optical Transceivers \vspace{-0.cm}} \author{Agust\'in
  C. Galetto, Benjam\'in T. Reyes, Dami\'an A. Morero,
  and~Mario~R.~Hueda%
  \thanks{Agust\'in C. Galetto and Benjam\'in T. Reyes are with
    Fundaci\'on Fulgor - Romagosa 518 - C\'ordoba (5000) -
    Argentina.}%
  \thanks{Dami\'an A. Morero and Mario R. Hueda are with Laboratorio
    de Comunicaciones Digitales - Universidad Nacional de C\'ordoba -
    CONICET - Av. V\'elez Sarsfield 1611 - C\'ordoba (X5016GCA) -
    Argentina.}  \vspace{.0cm}}

\markboth{}
{A.~GALETTO \MakeLowercase{\textit{et al.}}: Adaptive Background
  Compensation Technique for FI DACs with Application to Coherent
  Optical Transceivers}
\maketitle

\begin{abstract}
  This work proposes a novel adaptive background compensation scheme
  for frequency interleaved digital-to-analog converters
  (FI-DACs). The technique is applicable to high speed digital
  transceivers such as those used in coherent optical
  communications. Although compensation of FI-DACs has been discussed
  before in the technical literature, adaptive background techniques
  have not yet been reported. The importance of the latter lies in
  their capability to automatically compensate errors caused by
  process, voltage, and temperature variations in the technology
  (e.g., CMOS, SiGe, etc.) implementations of the data converters, and
  therefore ensure high manufacturing yield. The key ingredients of
  the proposed technique are a multiple-input multiple-output~(MIMO)
  equalizer and the \textit{backpropagation} algorithm used to adapt
  the coefficients of the aforementioned equalizer. Simulations show
  that the impairments of the analog signal path are accurately
  compensated and their effect is essentially eliminated, resulting in
  a high performance transmitter system.

\end{abstract}

\begin{IEEEkeywords}
  Frequency interleaving DAC, high-speed optical transmitter,
  background calibration, error backpropagation.\vspace{-0.cm}
\end{IEEEkeywords}

\IEEEpeerreviewmaketitle

\section{Introduction}\label{Sec:Introduction}

\IEEEPARstart{I}{ntensity} modulation and direct detection in long
haul and metro optical fiber links have been displaced by coherent
transmission techniques~\cite{morero_design_2016,
  faruk_digital_2017}. Next generation coherent transceivers will
operate at symbol rates of $f_B=$128-150 GBd and beyond
\cite{nakamura_advanced_2018}. The main challenge in the design of
transceivers for high-speed optical communications is achieving the
large bandwidth (BW) and sampling rate required by the
analog-to-digital and the digital-to-analog converters~(ADC and
DAC). One of the solutions proposed in recent literature is the use of
\emph{frequency interleaving} (FI) techniques
\cite{pupalaikis_technologies_2014, nakamura_104_2019}.

This paper focuses on FI-DACs in the context of applications to
transceivers for digital communications, particularly high baud rate
coherent optical transceivers. Although laboratory experiments with
high baud rate optical transmission based on FI-DAC have been
described in the technical literature \cite{chen_generation_2018},
significant obstacles remain before this technology can be applied in
commercial products. One of the main challenges is how to
automatically compensate the impairments of the analog signal path. As
it shall be discussed later, because of process tolerances, layout
limitations, etc., errors may exist which, if left uncompensated,
would introduce large distortions and severely degrade the performance
of the system. Several techniques have been presented to compensate
analog errors in FI architectures. In previous
work~\cite{schmidtinterleaving}, the compensation requires startup
calibration which is done using foreground techniques. In coherent
optical systems, the latter would imply the interruption of the
communication to compensate the imperfections, which is
undesirable. The compensation needs to be accurately tailored to the
impairments, which are process, voltage, and temperature dependent and
(slowly) time variant. The only way to achieve this at low cost and in
a way that lends itself to high volume manufacturing is to use
adaptive background compensation techniques. However, no adaptive
background compensation techniques for FI-DACs have been reported so
far in the technical literature.

\begin{figure}[!t]
  \centering
  \resizebox*{1.\columnwidth}{!}{\includegraphics{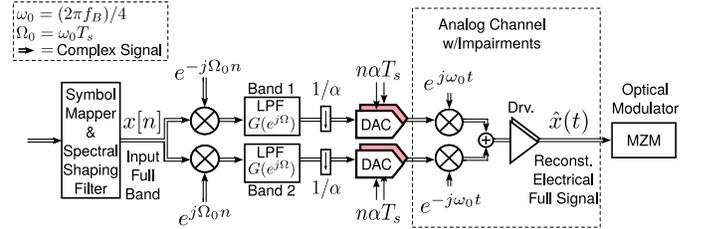}}
  \caption{\label{fig:fig1_model} Block diagram of a two-band FI-DAC
    architecture for DP optical coherent transceiver (only one
    polarization is depicted).}
  \vspace{-.20cm}
\end{figure}
\begin{figure}[!t]
  \centering
  \resizebox*{0.8\columnwidth}{!}{\includegraphics{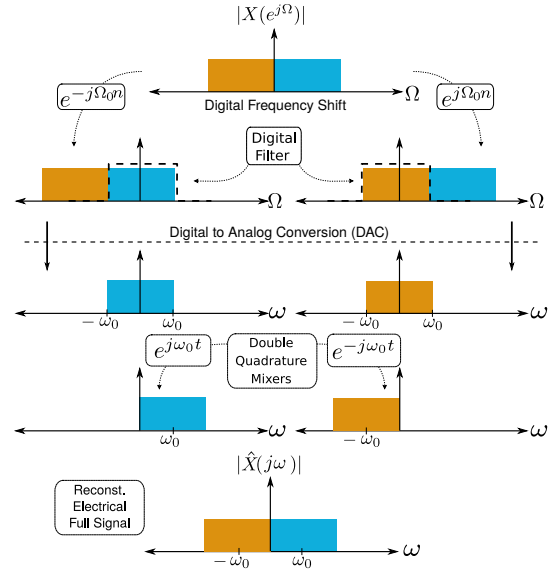}}
  \caption{\label{fig:fig0_bands} Example of the spectrums in a two-band FI-DAC architecture.}
  \vspace{-.20cm}
\end{figure}

The FI-DAC architecture considered in this paper is shown in
Fig. \ref{fig:fig1_model}. It is discussed in the context of its
application to a high baud rate transmitter for coherent optical
communications. The scheme of Fig. \ref{fig:fig1_model} corresponds to
one polarization in a dual-polarization (DP) coherent optical
transceiver. The transmit path is partitioned into two or more bands
by DSP techniques. Without loss of generality, in this paper we assume
that it is decomposed into two bands. As shown in Fig. \ref{fig:fig0_bands}, each band is demodulated to
baseband with two exponentials $e^{\pm j\Omega_0 n}$ where
$\Omega_0=\omega_0 T_s$ with $\omega_0=(2\pi f_B)/4$ and $1/T_s$ being
the DSP sampling rate. The demodulated baseband signals are first
processed by lowpass filters (LPF) with frequency response
$G(e^{j\Omega})$, then they are downsampled by a factor $\alpha$, and
synthesized by DACs of lower bandwidth and sampling rate than required
by the full signal. After synthesis, analog double quadrature mixers
reconstruct the full signal, which, after amplification by a modulator
driver, is used to control the Mach-Zehnder Modulator
(MZM)\footnote{{As discussed in Section~\ref{Sec:FI_DAC}, the analog
    path compensated by the scheme proposed here includes the
    impairments of the modulator driver and the interconnections among
    the FI-DAC, the driver, and the MZM. Said analog path may
    encompass components in different packages and printed circuit
    board (PCB) interconnects.}}. The reference clocks of the mixers
and the DACs are assumed to be properly synchronized. To process even
larger BW signals, this technique can be extended to more subbands,
which would require more DACs with similar characteristics as those
just described.

\begin{figure}[!t]
  \centering
  \resizebox*{1.\columnwidth}{!}{\includegraphics{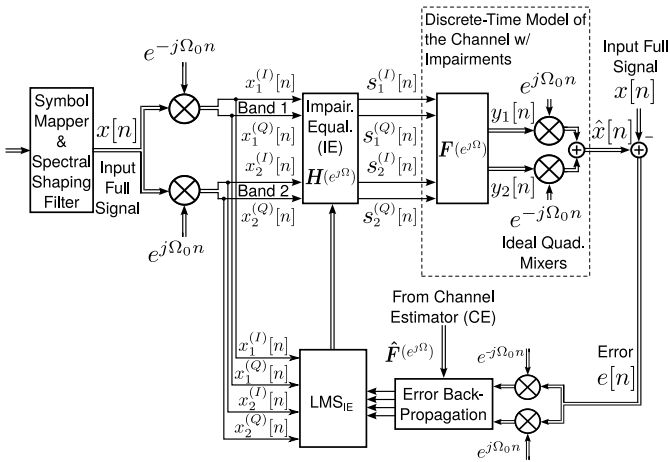}}
  \vspace{-0.cm} \caption{\label{fig:back_error} Proposed impairment
    equalizer (IE) and equivalent discrete time channel model of the
    analog impairments in a two-band FI-DAC based transmitter.}
  \vspace{-.20cm}
\end{figure} 

In this work we show that all analog impairments in a two-band FI-DAC
can be modeled as a $4\times 4$ multiple-input multiple-output~(MIMO)
real filter defined by a $4\times 4$ transfer matrix
$\mathbf{F}(j\omega)$, followed by ideal quadrature modulators (see
Section \ref{Sec:FI_DAC}). Fig. \ref{fig:back_error} depicts a block
diagram of the proposed compensation architecture and an equivalent
discrete time model of the two-band FI-DAC based coherent optical
transmitter. To compensate the effects of the impairments
$\mathbf{F}(e^{j\Omega})$, we introduce a MIMO adaptive equalizer
called hereafter \emph{Impairment Equalizer}~(IE) and defined by the
transfer matrix $\mathbf{H}(e^{j\Omega})$, which includes the LPF
responses $G(e^{j\Omega})$. Let $e[n]=\hat x[n]-x[n]$ be the error
between the reconstructed ($\hat x[n]$) and the original wideband
($x[n]$) full signal. This error is measured at the input of the MZM
(note that $\hat x[n]$ includes all the impairments of the analog path
up to the input of the MZM). Then, the IE is adapted to minimize the
mean squared error (i.e., $E\{|e[n]|^2\}$) by using the least mean
square (LMS) algorithm. Towards this end, the digital
\emph{backpropagation algorithm}~\cite{rumelhart_learning_1986,
  goodfellow_deep_2016} is proposed to perform background
compensation of the channel impairments\footnote{Alternatively, the
  \emph{forward propagation algorithm} \cite{morero_forward_2018}
  could be used. This option will be described in detail in a future
  paper.}. This algorithm, in combination with the estimated channel
response $\mathbf{\hat F}(e^{j\Omega})$, provides the error samples
required by the LMS algorithm to adapt the coefficients of the
IE. Computer simulations demonstrate that the proposed IE architecture
is able to compensate not only DACs and mixer impairments, but also
the amplitude and phase distortions of the electrical paths (e.g., it
acts as pre-emphasis and/or compensator of time skew between in-phase
and quadrature (I\&Q) components).

The rest of this paper is organized as
follows. Section~\ref{Sec:FI_DAC} introduces a model of the channel
impairments in FI-DACs for a coherent optical transceiver.
Section~\ref{Sec:backandforwardpropagation} describes the proposed
adaptive background compensation technique.  Section~\ref{Sec:Results}
presents simulations and finally conclusions are drawn in
Section~\ref{Sec:Conclusion}.

\section{Model of Analog Impairments in FI-DACs}\label{Sec:FI_DAC}

\begin{figure}[!t]
  \centering
  \resizebox*{.95\columnwidth}{!}{\includegraphics{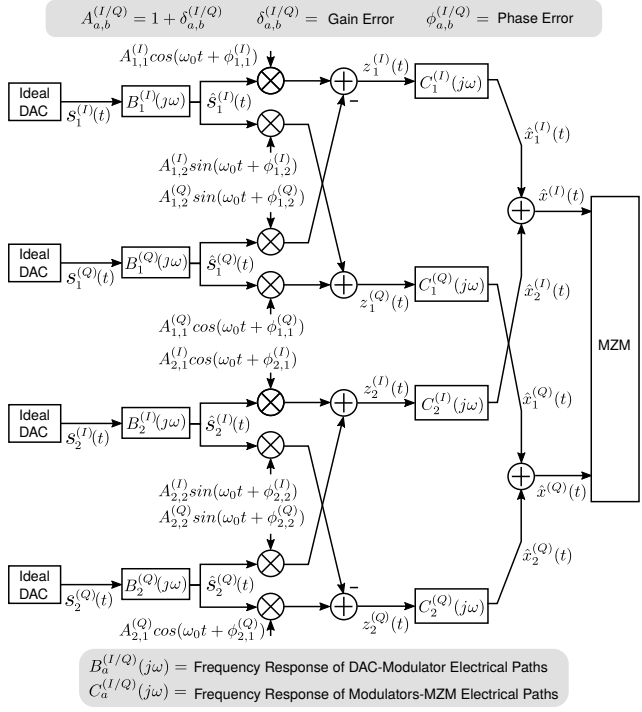}}
  \vspace{-0.cm} \caption{\label{fig:impair_model} Analog impairments
    in a two-band FI-DAC based DP coherent optical transceiver (only
    one polarization is depicted).}
\vspace{-0.2cm}
\end{figure}

Analog impairments drastically affect the performance of any FI-DAC
architecture, including the one presented here.  The most important
ones for a two-band FI-DAC based optical coherent transmitter are
shown in Fig. \ref{fig:impair_model} and include: \emph{(i)}
distortion, bandwidth limitation, and (in-band) time skew
($\tau_{IB}$) caused by mismatches among electrical path responses
between DACs and the quadrature mixers (denoted as
$B_a^{(I/Q)}(j\omega)$ with $a\in\{1,2\}$); \emph{(ii)}~gain and phase
errors (denoted, respectively, as $\delta^{(I/Q)}_{a,b}$ and
$\phi^{(I/Q)}_{a,b}$ with $a,b\in \{1,2\}$) of quadrature mixers
employed for the reconstruction of the full analog signal;
\emph{(iii)} distortion, bandwidth limitations, and time skew ($\tau$)
caused by mismatches among the electrical path responses going from
the quadrature mixers to the optical modulator (denoted as
$C_a^{(I/Q)}(j\omega)$ with $a\in\{1,2\}$ which include the modulator
drivers and any other components in the signal path). The effect of
the phase and gain errors in the quadrature modulators is to create
spurious terms that cause interference in the adjacent band.  As we
shall show later, this interference as well as the other impairments
can be compensated by the technique proposed in this paper.

\begin{figure}[!t]
  \centering
  \resizebox*{1.\columnwidth}{!}{\includegraphics{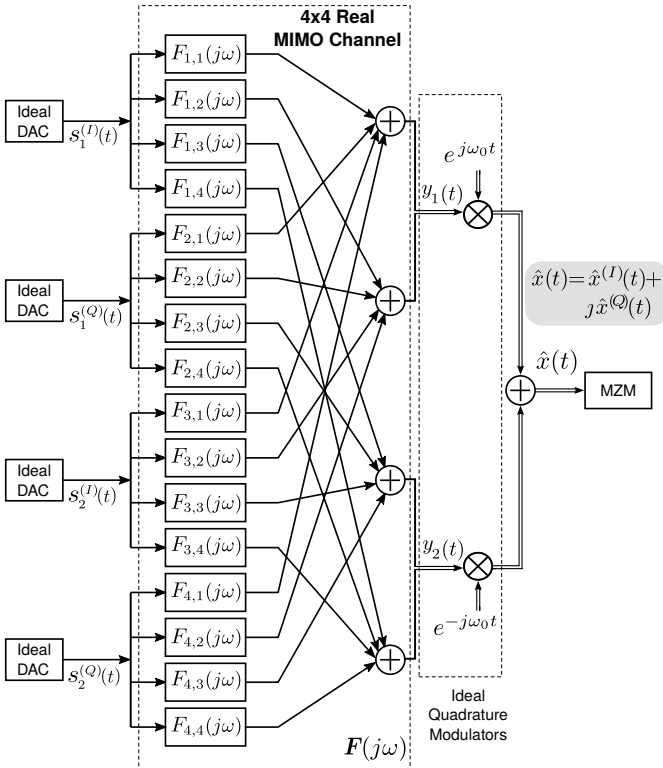}}
  \vspace{-0.cm} \caption{\label{fig:eq_impair_model} Equivalent
    channel model of analog impairments in a two-band FI-DAC for one
    polarization in a DP coherent optical transmitter.}
\vspace{-0.cm}
\end{figure}

Based on simple trigonometric identities and signal processing
techniques, we show in the Appendix that the analog channel model with
impairments for one polarization described in
Fig. \ref{fig:impair_model} can be reformulated as a $4\times 4$ MIMO
real channel defined by a $4\times 4$ transfer matrix
$\mathbf{F}(j\omega)$ with elements $F_{u,v}(j\omega),$
$u,v\in\{1,...,4\}$, followed by ideal quadrature modulators (see
Fig. \ref{fig:eq_impair_model}). Based on this result, we can derive a
simple discrete time model of the FI-DAC architecture for application
in an optical coherent transmitter as depicted in
Fig. \ref{fig:back_error}. This formulation is important since it
shows that all the impairments in FI-DACs can be digitally compensated
by a MIMO compensator equalizer with transfer matrix
$\mathbf{H}(e^{j\Omega})$. For example, for an ideal compensation, we
get that
$\mathbf{H}(e^{j\Omega})\mathbf{F}(e^{j\Omega})=G(e^{j\Omega})\mathbf{I}_4$ where $\mathbf{I}_4$ is the $4\times 4$ identity matrix and
$G(e^{j\Omega})$ is the Fourier transform (FT) of the lowpass filter
impulse response depicted in Fig. \ref{fig:fig1_model}.

\begin{figure}[!t]
  \centering
  \resizebox*{1.\columnwidth}{!}{\includegraphics{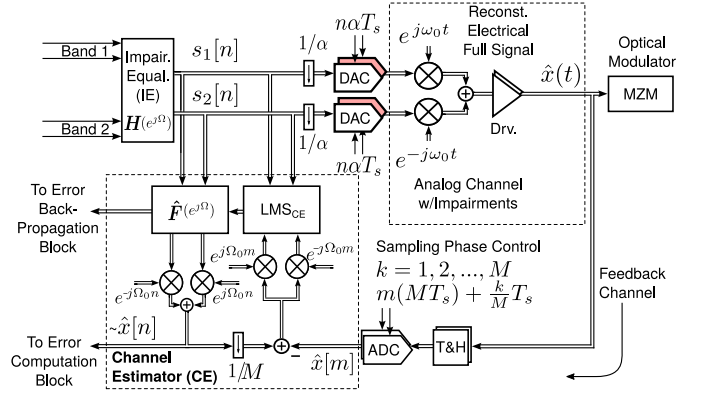}}
  \vspace{-0.cm} \caption{\label{fig:feedback_model} Low complexity
    architecture for estimating the equivalent channel model of analog
    impairments in a two-band FI-DAC for one polarization in a DP
    coherent optical transmitter.}
\vspace{-0.cm}
\end{figure}

\subsection{Channel Estimator (CE) Block}\label{subsec:DSP_Based_Compensation}
As we shall discuss in the next section, the proposed compensation
scheme is based on the evaluation of the error $e[n]=\hat x[n]-x[n]$
and the background estimation of the analog path response with
impairments, $\mathbf{F}(e^{j\Omega})$. These operations are based on
the samples of the output signal $\hat x(t)$ as depicted in
Fig. \ref{fig:feedback_model}. The feedback path includes buffers and
track and holds (T\&H) to support the bandwidth of the reconstructed
full signal $\hat x(t)$. However, since channel impairments change
slowly over time, the estimation algorithm does not need to operate at
full rate. Therefore, low power, low speed (i.e., $1/(M T_s)$ with
$M\gg 1$), medium resolution ADCs with adjustable sampling phase can
be used. The estimation of the analog channel response
$\mathbf{F}(e^{j\Omega})$ can be achieved by using the LMS algorithm
(LMS$_{CE}$) and the error between the DAC inputs (e.g., $s_1[n]$ and
$s_2[n]$) and the samples of the reconstructed full signal as depicted
in Fig. \ref{fig:feedback_model}. Notice that CE block is also able to
provide samples of the reconstructed signal at full rate for
computation of the error $e[n]$. The response of the feedback path
(i.e., T\&H, ADC) could be initially estimated and removed from
$\mathbf{\hat F}(e^{j\Omega})$ (the details are omitted due to space
constraints).

\section{Adaptation of the Impairment Equalizer (IE): Error
  Backpropagation Algorithm}\label{Sec:backandforwardpropagation}

The samples of the reconstructed full signal for one polarization can
be expressed as (see Figs. \ref{fig:back_error} and
\ref{fig:eq_impair_model}):
\begin{equation}
  \label{eq:hatxn}
  \hat x[n]= y_1[n]e^{j\Omega_0 n}+y_2[n]e^{-j\Omega_0 n},
\end{equation}
where $y_1[n]=y^{(I)}_1[n]+jy^{(Q)}_1[n]$ and
$y_2[n]=y^{(I)}_2[n]+jy^{(Q)}_2[n]$ with components given by
\begin{equation}
  \label{eq:yn}
  \mathbf{y}[n]=\mathcal{F}^{-1} \lbrace \mathbf{F}(e^{j\Omega})
  \mathcal{F}\lbrace\mathbf{s}[n]\rbrace \rbrace,
\end{equation}
where $\mathcal{F}\lbrace.\rbrace$ ($\mathcal{F}^{-1}\lbrace.\rbrace$)
denotes the FT (inverse FT) operator, $\mathbf{y}[n]$ is the
$4\times 1$ real vector defined by
$\mathbf{y}[n]=[y^{(I)}_1[n] \; y^{(Q)}_1[n] \; y^{(I)}_2[n] \;
y^{(Q)}_2[n]]^{T}$ while
$\mathbf{s}[n]=[s^{(I)}_1[n] \; s^{(Q)}_1[n] \; s^{(I)}_2[n] \;
s^{(Q)}_2[n]]^{T}$ is the $4\times 1$ real vector with the digital
samples at the DAC inputs.

Let ${H}_{u,v}(e^{j\Omega})$ with $u,v\in\{1,2,3,4\}$ be the $(u,v)$
transfer function of the $4 \times 4$ transfer matrix of the IE,
${\bold H}(e^{j\Omega})$. We also define the \emph{real} impulse
responses
$h_{u,v}[n]=\mathcal{F}^{-1}\lbrace {H}_{u,v}(e^{j\Omega})\rbrace$
with $u,v\in\{1,2,3,4\}$. In this work we adopt the LMS algorithm to
iteratively adapt the real coefficients of the set $h_{u,v}[n]$ in
order to minimize the mean squared error (MSE) between the input and
the reconstructed full samples (LMS$_{IE}$):
\begin{equation}
  \label{eq:hn}
  {\bold h}^{(k+1)}_{u,v}={\bold h}^{(k)}_{u,v}- \beta
  \nabla_{{\bold h}_{u,v}} E\{|e[n]|^2\},\quad u,v\in\{1,...,4\},
\end{equation}
where $k$ denotes the number of iteration,
${\bold h}_{u,v}=[h_{u,v}[0]\;
h_{u,v}[1]\;\cdots\;h_{u,v}[L_h-1]\;]^T$, $L_h$ is the number of
coefficients of the filters, $\beta$ is the adaptation step, and
$\nabla_{{\bold h}_{u,v}} E\{|e[n]|^2\}$ is the gradient of the MSE
with respect to the real vector ${\bold h}_{u,v}$. The computation of
the latter is not trivial since $e[n]$ is not the error at the output
of the IE (see Fig. \ref{fig:back_error}).

To get the proper error samples to adapt the coefficients of the
filters as expressed in eq. \eqref{eq:hn}, we propose to use the
\emph{backpropagation algorithm} widely used in \emph{machine
  learning} \cite{rumelhart_learning_1986,
  goodfellow_deep_2016}. Towards this end, we first generate the
demodulated band errors $e_1[n]=e[n]e^{j\Omega_0 n}$ and
$e_2[n]=e[n]e^{-j\Omega_0 n}$. These errors, in combination with
eq. \eqref{eq:yn} and the estimated channel response
$\mathbf{\hat F}(e^{j\Omega})$, are used to get the backpropagated
errors $\tilde e_1[n]$ and $\tilde e_2[n]$
(see~\cite{morero_forward_2018,rumelhart_learning_1986,
  goodfellow_deep_2016} for a detailed description of the
backpropagation technique). Finally, based on these backpropagated
errors we can estimate the gradient
$\nabla_{{\bold h}_{u,v}} E\{|e[n]|^2\}$ as usual in the classical LMS
algorithm.

Since channel impairments change slowly over time, the coefficient
adaptation does not need to operate at full rate, and subsampling can
be applied. The latter allows implementation complexity to be
significantly reduced. Additional complexity reduction is enabled by:
1) strobing the algorithms once they have converged, and/or 2)
implementing them in firmware in an embedded processor, typically
available in coherent optical transceivers.

\vspace{-0.2cm}
\section{Simulations}\label{Sec:Results}
We investigate the performance of the proposed background calibration
technique in a two-band FI-DAC based DP coherent optical system by
using computer simulations. We assume 16-QAM modulation with a symbol
rate of $f_B=1/T=128$ GBd in a back-to-back optical channel. The
oversampling factor used in the DSP blocks is $T/T_s=2$. We consider
8-bit resolution DACs with sampling rate of 128 GS/s (i.e., $\alpha=2$
in Fig. \ref{fig:fig1_model}) and nominal BW of $B_0=$32 GHz, which is
half of what would be needed to process the input signal band in a
non-interleaved architecture. The electrical analog path responses
$B_a^{(I/Q)}(j\omega)$ with $a\in\{1,2\}$ in
Fig. \ref{fig:impair_model} are simulated with third-order Butterworth
lowpass filters with nominal BW $B_0$. Ideal
feedback channel is assumed. The number of taps of the impairment
equalizers ${\bold h}_{u,v}$ is $L_h=21$. The subsampling factor of
the feedback ADC is $M=128$. Other details of the DSP blocks are
omitted due to space limitations (see \cite{morero_design_2016} and
references therein for details of typical DSP blocks).
\begin{figure}[!t]
  \centering
  \resizebox*{.98\columnwidth}{!}{\includegraphics{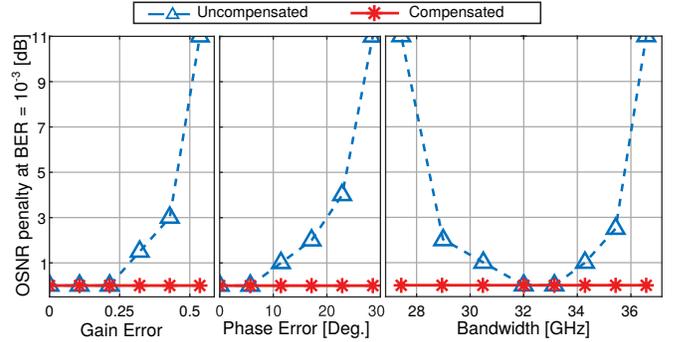}}
  \vspace{-0.cm} \caption{\label{fig:results} OSNR penalty at
    BER$=10^{-3}$ versus gain error, phase error, and bandwidth
    mismatches of the electrical paths $B_a^{(I/Q)}(j\omega)$ (see
    Fig. \ref{fig:impair_model}).}
  \vspace{-.2cm}
\end{figure} 
We focus on the optical signal-to-noise ratio (OSNR)
penalty\footnote{See \cite{chan_optical_2010} for a definition of OSNR
  penalty.} at a bit-error-rate (BER) of $10^{-3}$, which is computed
using an ideal software coherent receiver.

Fig. \ref{fig:results} shows the OSNR penalty as a function of the
gain and phase errors of the quadrature mixers, and the bandwidth
mismatches, respectively. Only one effect is exercised in each
case. To stress the mismatch effects, the impairments are added only
to the paths of signals $s_1^{(Q)}(t)$ and $s_2^{(I)}(t)$ of each
polarization (see Fig. \ref{fig:impair_model}). We present results
with and without the proposed background compensation technique. The
effectiveness of the IE architecture is verified in all cases. Notice
that the proposed FI-DAC scheme performs very well in the presence of
inaccuracies of the analog BW.
\begin{figure}[!t]
  \centering
  \resizebox*{.6\columnwidth}{!}{\includegraphics{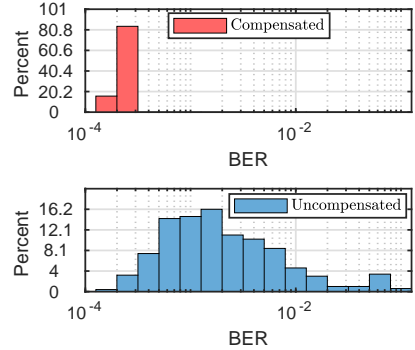}}
  \vspace{-0.cm} \caption{\label{fig:results2} Histogram of BER for
    500 random analog channels with and without compensation for a BER
    reference of $\sim 2\times 10^{-4}$.}
  \vspace{-.3cm}
\end{figure} 

The robustness of the proposed compensation scheme was assessed by
running Montecarlo simulations. A total of 500 channels with
uniformely distributed random impairments such as gain errors
($\delta^{(I/Q)}_{a,b}\in [\pm 0.25]$), phase errors
($\phi^{(I/Q)}_{a,b}\in [\pm 22^o]$), and bandwidth mismatches
($\Delta B_0/B_0 \in [\pm 25\%]$), were simulated. We set the OSNR to
that required to achieve a BER $\sim 2\times 10^{-4}$ in the absence
of analog impairments. Fig.\ref{fig:results2} shows the histogram of
the BER for the system with and without the MIMO equalizer. In all
cases the excellent compensation of the impairments achieved by the
proposed IE architecture is verified.

\vspace{-.0cm}
\section{Conclusions}\label{Sec:Conclusion}
An FI-DAC architecture with adaptive background compensation of the
analog signal path errors for coherent optical transceivers has been
presented. Simulations show the effectiveness of the proposed
technique, which results in the elimination of the penalty caused by
the DAC frequency response and the gain and phase errors in the
mixers, as well as other impairments. Although the technique was
presented in the context of a transmitter for coherent optical
communications, it is more general and can be used in other
applications.

\vspace{-.0cm}
\section*{Appendix}\label{Sec:Appendix} In this Appendix
we derive the channel model of Fig. \ref{fig:eq_impair_model}. With a
proper design, it is possible to assume that the two cosine (and sine)
signals used in each quadrature mixer have the same gain and
phase\footnote{Although this assumption simplifies the math, it can be
  shown that it is not necessary for the applicability of the
  compensation scheme proposed in this paper.}, i.e.,
$A_{a,1}^{(I)}=A_{a,1}^{(Q)}=1+\delta_a$,
$A_{a,2}^{(I)}=A_{a,2}^{(Q)}=1-\delta_a$,
$\phi_{a,1}^{(I)}=\phi_{a,1}^{(Q)}=\phi_a/2$, and
$\phi_{a,2}^{(I)}=\phi_{a,2}^{(Q)}=-\phi_a/2$ with $a\in\{1,2\}$. Note
that $\delta_a$ and $\phi_a$ represent the gain and phase imbalance of
the quadrature mixer for band $a$, respectively. Then, based on the
complex model proposed in \cite{daSilva16}, the modulator output of
band $a$ denoted as $z_a(t)=z_a^{(I)}(t)+jz_a^{(Q)}(t)$ (see
Fig. \ref{fig:impair_model}), can be expressed as
\begin{equation}
  \label{eq:za}
  z_a(t)= \hat s_a(t)p_a(t),\quad a\in\{1,2\},
\end{equation}
where $\hat s_a(t)=\hat s_a^{(I)}(t)+j\hat s_a^{(Q)}(t)$ is the mixer
input, and
\begin{equation}
  \label{eq:pa}
  p_a(t)= k_{a,1}e^{j\omega_0 t}+k_{a,2}e^{-j\omega_0 t}
\end{equation}
with complex constants given by
\begin{align}
  \label{eq:k11}
  k_{1,1}&= k^{(I)}_{1,1}+jk^{(Q)}_{1,1}=\frac{1+\delta_1}{2}e^{j\phi_1/2}+\frac{1-\delta_1}{2}e^{-j\phi_1/2},\\
  \label{eq:k12}
  k_{1,2}&= k^{(I)}_{1,2}+jk^{(Q)}_{1,2}=\frac{1+\delta_1}{2}e^{-j\phi_1/2}-\frac{1-\delta_1}{2}e^{j\phi_1/2},\\
  \label{eq:k21}
  k_{2,1}&= k^{(I)}_{2,1}+jk^{(Q)}_{2,1}=\frac{1+\delta_2}{2}e^{j\phi_2/2}-\frac{1-\delta_2}{2}e^{-j\phi_2/2},\\
  \label{eq:k22}
  k_{2,2}&= k^{(I)}_{2,2}+jk^{(Q)}_{2,2}=\frac{1+\delta_2}{2}e^{-j\phi_2/2}+\frac{1-\delta_2}{2}e^{j\phi_2/2}.
\end{align}

\begin{figure}[!t]
  \vspace{-0.cm} \centering
  \resizebox*{1.\columnwidth}{!}{\includegraphics{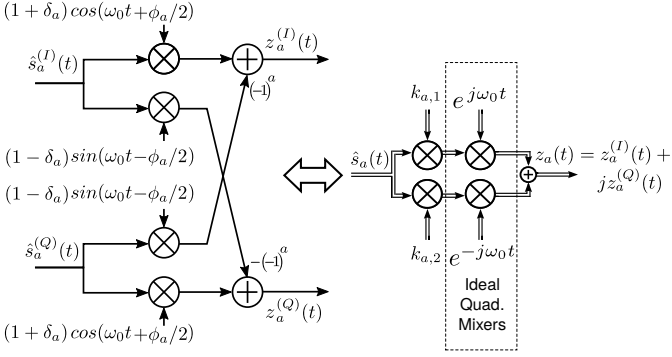}}
  \vspace{-0.3cm} \caption{\label{fig:mixer_model} Equivalent complex
    channel model of a quadrature mixer with I/Q imbalance for a the
    subband $a\in\{1,2\}$.}
\vspace{-0.cm}
\end{figure}

In the absence of gain and phase errors, notice that
$p_1(t)= e^{j\omega_0 t}$ and $p_2(t)= e^{-j\omega_0
  t}$. Fig. \ref{fig:mixer_model} shows the equivalent complex-valued
model of the quadrature mixer with impairments.
\begin{figure}[!t]
  \centering
  \resizebox*{1.\columnwidth}{!}{\includegraphics{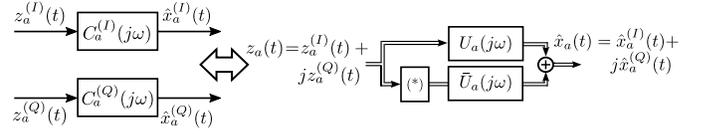}}
  \vspace{-0.2cm} \caption{\label{fig:ele_path_model} Complex-valued
    channel model of the mixer-MZM electrical paths with responses
    $C_a^{(I/Q)}(j\omega)$ for the subband $a\in\{1,2\}$ (see
    Fig. \ref{fig:impair_model}).}  \vspace{-0.cm}
\end{figure} 
The quadrature mixer outputs $z^{(I/Q)}_a(t)$ are transmitted up to
the optical modulator through electrical paths (which include the
modulator drivers) with mismatch responses $C_a^{(I/Q)}(j\omega)$ (see
Fig. \ref{fig:impair_model}). From eq. \eqref{eq:za} and
Fig. \ref{fig:ele_path_model}, the complex signal at the MZM input can
be expressed as
\begin{equation}
  \label{eq:xa}
  \hat x_a(t)= z_a^{(I)}(t)\otimes c^{(I)}_a(t)+jz_a^{(Q)}(t)\otimes c^{(Q)}_a(t),
\end{equation}
where symbol $\otimes$ denotes the convolution operation and
$c^{(I/Q)}_a(t)=\mathcal{F}^{-1}\lbrace
C_a^{(I/Q)}(j\omega)\rbrace$. Since
$z_a^{(I)}(t)=0.5[z_a(t)+z^{*}_a(t)]$ and
$jz_a^{(Q)}(t)=0.5[z_a(t)-z^{*}_a(t)]$, we can get
\begin{equation}
  \label{eq:xab}
  \hat x_a(t)= z_a(t)\otimes u_a(t)+z_a^{*}(t)\otimes {\overline u}_a(t),
\end{equation}
where $u_a(t)=0.5[c^{(I)}_a(t)+c^{(Q)}_a(t)]$ and
${\overline u}_a(t)=0.5[c^{(I)}_a(t)-c^{(Q)}_a(t)]$ (see
Fig. \eqref{fig:ele_path_model}).

\begin{figure}[!t]
  \centering
  \resizebox*{1.\columnwidth}{!}{\includegraphics{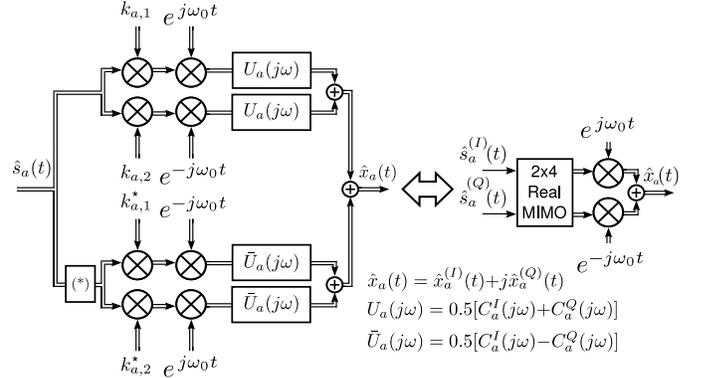}}
  \vspace{-0.4cm} \caption{\label{fig:eq_ep_model} Equivalent MIMO
    channel model of the combined quadrature mixer and mixer-MZM
    electrical paths for the subband $a\in\{1,2\}$.}
  \vspace{-0.2cm}
\end{figure}

From the above, we can derive the channel model for the quadrature
mixer with impairments including the electrical paths as shown in the
block diagram on the left side of
Fig. \ref{fig:eq_ep_model}. Moreover, taking into account that
$\hat s_a(t)e^{\pm j\omega_0 t}\otimes u(t)=[\hat s_a(t)\otimes
u(t)e^{\mp j\omega_0 t}]e^{\pm j\omega_0 t}$, it is possible to
exchange the order of the modulator and filter blocks ${u}_a(t)$ and
${\overline u}_a(t)$. Then, grouping the signals properly, the analog
channel for the subband $a$ can be reduced to a $2\times 4$ MIMO real
channel followed by two ideal quadrature mixers as depicted on the
right side of Fig. \ref{fig:eq_ep_model}. Notice that the response of
the DAC-mixer electrical paths (i.e., $B_a^{(I/Q)}(j\omega)$ in
Fig. \ref{fig:impair_model}) can be easily included within this MIMO
channel model. Finally, the model just described is applied to the two
subbands which are combined resulting in the $4\times 4$ MIMO real
channel model of Fig. \ref{fig:eq_impair_model}.

\bibliographystyle{./IEEEtran/bibtex/IEEEtran} 
\bibliography{./IEEEtran/bibtex/IEEEabrv.bib,bib_paper.bib}

\end{document}